\documentclass[10pt]{article}
\usepackage{epsfig}
\usepackage{amssymb}
\usepackage{amsmath}
\newcommand{\ud}{\mathrm{d}}

\begin{document}


\title{\bf SOLITONIC AND NON-SOLITONIC Q-STARS\thanks{Talk given at the
 11th Marcel Grossmann Meeting, Berlin, 23-29 July 2006.}}

   \author{
  \large Y. Verbin }
 \date{ }
   \maketitle
       \centerline{\em Department of Natural Sciences, The Open University
   of Israel,}
   \centerline{\em P.O.B. 39328, Tel Aviv 61392, Israel}
  \centerline{ e-mail: \em verbin@openu.ac.il}
   \vskip 1.1cm

   \begin{abstract}
The properties of several types of Q-stars are studied and compared with their flat 
space analogues, i.e. Q-balls. The analysis is based on calculating the mass, global 
U(1) charge and binding energy for families of solutions parametrized by the central 
value of the  scalar field. The two most frequently used Q-star models (differing by 
their potential term) are studied. Although there are general similarities 
between both Q-star types, there are important differences as well as new features with 
respect to the non-gravitating systems. We find non-solitonic solutions which do not 
have a flat space limit, in the weak (scalar) field region as well as in the opposite 
region of strong central scalar field for which there does not exist Q-ball solutions 
at all. \\
 \end{abstract}

\setcounter{equation}{0}

Q-balls \cite{Coleman1985} are a simple kind of non-topological
solitons which occur in a wide variety of (theoretical) physical contexts 
\cite{Kusenko1997b,EnqvistMacD1998,KusenkoShaposh1998,EnqvistMacD1999,Kusenko1997a,
DvaliEtAl1997,EnqvistMazumdar2003,DineKusenko2004} like the 
 supersymmetric Standard Model \cite{Kusenko1997b,EnqvistMacD1998}.
 
Most of the Q-ball studies are based on the ``original'' flat space Q-balls. 
However, it is evident that for a large enough mass scale, gravitational effects become 
important and one needs to study Q-stars \cite{Lynn89} which are their self-gravitating
generalizations. That is, they are finite mass and charge solutions of the following 
U(1) symmetric action:
\begin{eqnarray}
S=\int \ud^4x\sqrt{|g|}\left( \frac{1}{2}(\nabla_\mu \Phi)^*(\nabla ^\mu
\Phi)-U(|\Phi|)+\frac{1}{16\pi {\cal G}}R\right)
\label{action1}
\end{eqnarray}
where the potential function is usually taken to be:
\begin{eqnarray}
U(|\Phi|)  = \frac{m^2}{2}|\Phi|^2-\frac{\alpha m^{4-p}}{p}|\Phi|^p+
\frac{\lambda m^{4-q}}{q}|\Phi|^q . \label{potentials}
\end{eqnarray}   
Two choices are popular in the literature: $p=3,q=4$ and $p=4,q=6$. 
Actually, this  system allows for a different kind of localized solutions already without 
self-interaction (i.e. only mass term) or with an additional $|\Phi|^4$ term, namely, boson 
stars \cite{Jetzer1992,LeePang1992,Liddle1992}. Boson stars have also a conserved
global U(1) charge, but unlike Q-stars, they do not have a flat space limit.

I will assume spherically-symmetric solutions with non-vanishing U(1) charge, i.e. 
$\Phi=mf(r)e^{i\omega t}$ and $ds^2=A^2(r)dt^2- B^2(r)dr^2-r^2(d\theta^2+\sin^2\theta d\varphi^2)$ 
so the charge and mass are given by
\begin{eqnarray}
Q=4\pi \omega m^2 \int_{0}^\infty dr r^2 (B/A) f^2
\,,\,\,\, M=4\pi  \int_{0}^\infty dr r^2
\left[\frac{\omega^2 m^2 f^2}{2A^2} +\frac{m^2 f'^2}{2B^2}+U(f)\right].
\label{Mass}
\end{eqnarray}
Without loss of generality we can assume $\omega>0$ so we will have $Q>0$ as well. 

The existence of Q-stars was demonstrated by Friedberg {\it et al} \cite{FriedbergLeePang1986}
and by Lynn \cite{Lynn89} together with a presentation of 
the basic properties of the solutions for the 2-4-6 potential. A discussion of 2-3-4 Q-stars 
 appeared only quite recently
 \cite{Prikas2002}. It was shortly followed by a study 
 \cite{MultamakiVilja2002} which showed that gravity limits the size of Q-balls. 
On the other hand, a recent analysis \cite{VolkovWohnert,KleihausEtAl2005} of 
spinning Q-balls and  Q-stars is concentrated in the 2-4-6 case. 

I give here the main results of a systematic comparative study of
both kinds of Q-stars, including the dependence on the gravitational 
 strength parametrized by the dimensionless parameter $\gamma=4\pi{\cal G}m^2$.  
 A more detailed summary will be presented elsewhere \cite{Verbin2006}.
 I choose in both cases the parameters $\alpha=2$ and $\lambda=1$ so the potentials 
 will have a similar form. The gravitational parameter $\gamma$  will take the 
 following three values: $\gamma=0,\,0.02,\, 0.2$.

\begin{figure}[!t]
      \begin{center}
       (a) \hspace{60mm} (b)\\
      \includegraphics[height=.2\textheight,width=.45\textwidth]{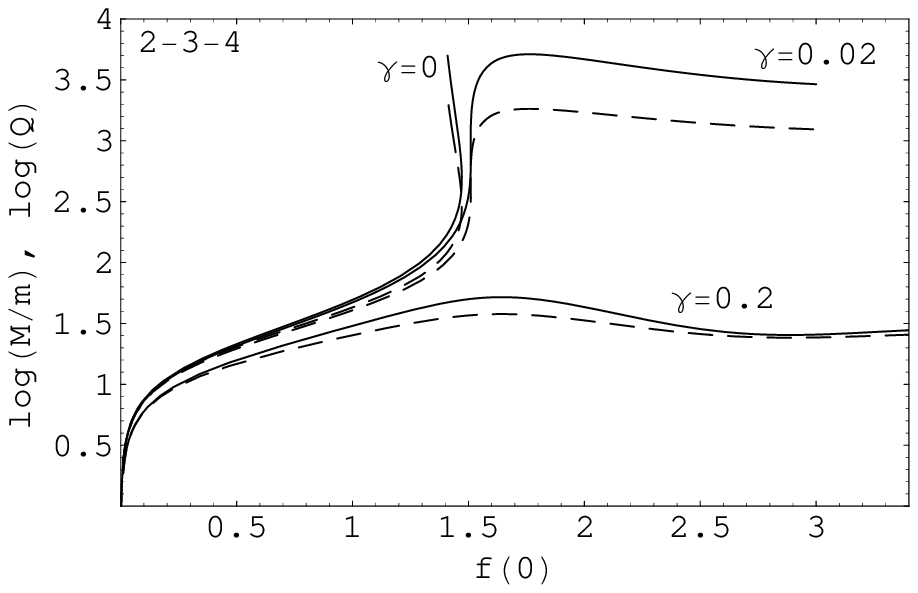} 
       \includegraphics[height=.2\textheight,width=.45\textwidth]{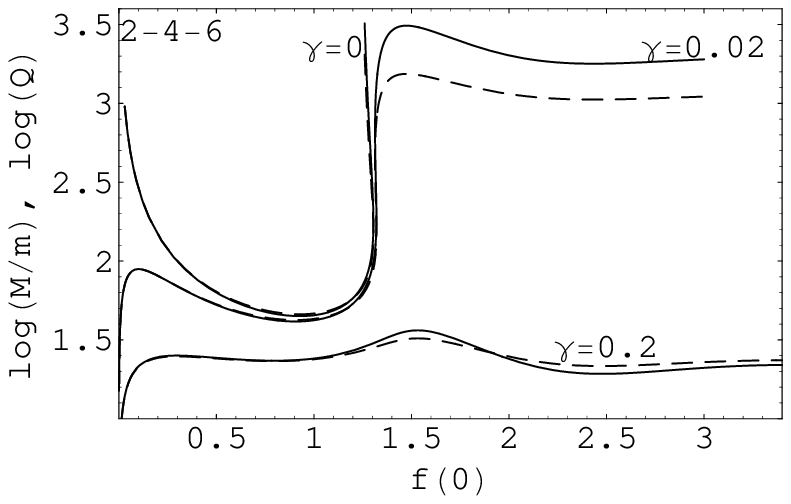}\\
   \caption{Plots of $\log(Q)$ (solid line) and $\log(M/m)$ (dashed) vs. 
   $f(0)$  for $\gamma=0$ (Q-balls),
    $\gamma=0.02$ and  $\gamma=0.2$. 
   (a) 2-3-4  Q-stars; (b) 2-4-6  Q-stars.}
 \label{figureQSLogQMfx0}
     \end{center}
     \end{figure}

\begin{figure}[!b]
   \begin{center}
     (a) \hspace{60mm} (b)\\
      \includegraphics[height=.2\textheight,width=.45\textwidth]{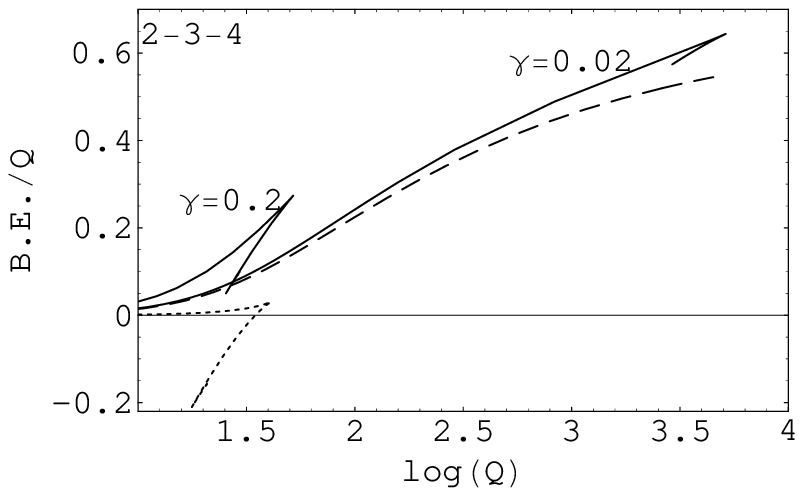} 
      \includegraphics[height=.2\textheight,width=.45\textwidth]{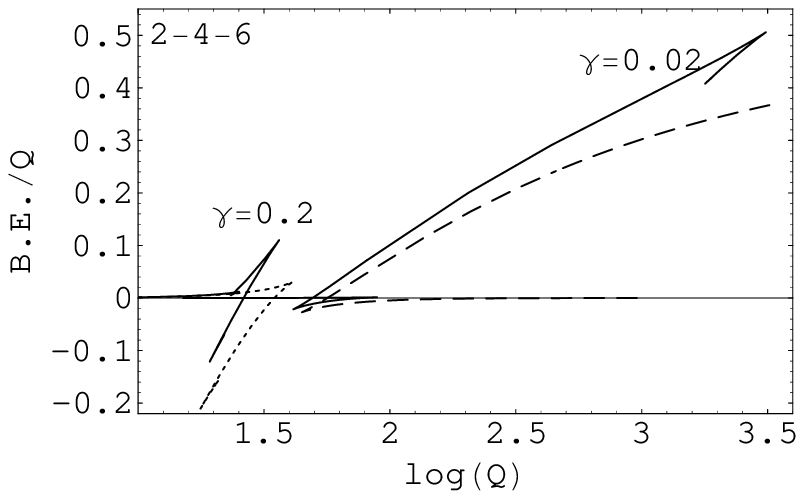}\\
   \caption{Plots of binding energy per particle $(mQ-M)/mQ$ vs. $\log(Q)$  for 
   $\gamma=0$ (dashed), $\gamma=0.02$ and  $\gamma=0.2$ and 
   for boson stars with $\gamma=0.2$ (dotted).  (a) 2-3-4  Q-stars; (b) 2-4-6  Q-stars.}
 \label{figureQSBEoverQvsLogQ}
     \end{center}
     \end{figure}

Figure \ref{figureQSLogQMfx0} summarizes the main results in the $Q-f(0)$ and $M-f(0)$ planes
namely, the general behavior of the charge and mass. Figure \ref{figureQSBEoverQvsLogQ} depicts the 
binding energy per particle (in dimensionless form), $1-M/mQ$ as a ``function'' of $Q$ which is
more instructive from a physical point of view. It is evident from this figure 
 that Q-stars are more strongly-bound than their non-gravitating counterparts with the same $Q$. 
From our results one can draw the following observations and conclusions: 

Already in flat space there is a very significant difference between the ``thick wall''  
(small $f(0)$) solutions of the two potentials: the thick wall Q-balls of the 2-3-4 potential 
are small and stable, i.e. $Q$ and $M$ vanish as
$\omega\rightarrow m$ or $f(0)\rightarrow 0$ while  $M/m<Q$. On the other
hand, those of the 2-4-6 potential are large and unstable, i.e. 
both charge and mass diverge while $M/m>Q$ in the same limit. 

Gravity introduces significant changes such as allowing solutions in regions where flat space 
solutions do not exist and limiting the charge and mass of Q-stars. But still the changes are 
quite small for weak scalar field 2-3-4 Q-stars, as seen in figures \ref{figureQSLogQMfx0}a
and \ref{figureQSBEoverQvsLogQ}a. On the other hand, gravity changes completely 
the nature of the weak field 2-4-6 solutions even for a small $\gamma$ (say, 0.02) 
as figure \ref{figureQSLogQMfx0}b shows: as $f(0)\rightarrow 0$ the charge and mass do not blow 
up, but on the contrary go to zero. Looking from the other direction, one sees that the charge and 
mass start rising from zero, reach a local maximum, decrease a little 
and then go to the thin wall region and beyond as described below. Moreover, unlike the 2-4-6 
Q-balls for $f(0)<<1$ which were unstable, now there appears a region of stability 
(below the resolution of figures \ref{figureQSLogQMfx0}b or \ref{figureQSBEoverQvsLogQ}b) 
for small enough $f(0)$. This is followed by a region of unstable 
solutions up to a certain value of $f(0)$. From this point on, all solutions are stable.

For larger values of $f(0)$ we encounter for both potentials ``thin wall'' Q-stars quite 
similar to the corresponding Q-balls, although the self-gravitating solutions are not so well 
described by the thin wall approximation. The reason is that where the 
thin wall approximation in flat space is accurate, gravity already causes deviations. 
To push it to the extreme, the thin wall approximation becomes exact for
 $Q\rightarrow\infty$, but gravity keeps Q-stars away from this  
 best region by introducing a maximal value of $Q$. For large values  
 of $\gamma$ there are no thin wall solutions altogether. 
 
 Another new gravitational effect is the existence of solutions when the central field becomes 
considerably larger than $f_{\ast}(0)$ which is the flat space critical field. 
Unlike the Q-ball case, the mass and charge curves cross this point and there are solutions 
as far as I was able to explore numerically. All small $\gamma$ solutions are stable, but their 
nature for $f(0)\gtrsim f_{\ast}(0)$ becomes 
quite different from the Q-balls as we go further.  Moreover, it is obvious that this kind of 
solutions cannot be 
considered solitonic as they do not have a flat space limit: while the charge and mass of 
the solutions with $f(0) < f_{\ast}(0)$ have a (finite) limit as $\gamma\rightarrow 0$, 
those in the other region blow up. In other words, it is only thanks to gravity 
that this kind of solutions with $f(0)\gtrsim f_{\ast}(0)$ exists.

\vfill

\end{document}